# Directional coupling and focusing surface plasmon polaritons controlled by arbitrary spin of light

HAILONG ZHOU[1,2], JINRAN QIE[1,2], JIANJI DONG[1,3] AND XINLIANG ZHANG[1,4]

[1]*Wuhan National Laboratory for Optoelectronics, School of Optical and Electronic Information, Huazhong University of Science and Technology, Wuhan, China, 430074*
[2]*These authors contributed equally to this work*
[3]*jjdong@hust.edu.cn*
[4]*xlzhang@mail.hust.edu.cn*

**Abstract:** Surface plasmon polaritons have attracted varies of interests due to its special properties, especially in the polarization-controlled devices. Typically, the polarization-controlled devices include directional coupling, focusing lens and plasmonic vortex lens, and almost all of them are controlled by the input circularly polarized light or the linearly polarized light. We present a novel device that realize the functions of directional coupling and focusing with high polarization extinction ratio for arbitrary spin of input light. This device offers opportunities for polarization sensing, polarization splitting and polarization-multiplexed near-field images and surface plasmon holography in the future.

## 1. Introduction

Surface plasmon polaritons (SPPs), propagating along and tightly bound to metal/dielectric interfaces, are electromagnetic excitations constituted by electromagnetic field in dielectric coupled to collective electron oscillations in metal [1]. It has become ubiquitous in varieties of areas, including miniaturized optoelectronic circuitry, light–matter interactions, high-resolution imaging, and ultrasensitive biochemical sensing [1, 2]. And in these applications, polarization-controlled devices based on SPPs has drawn an impressive interest during the past few years due to the strong independence of SPPs' properties on the composition and structure of the surface, which help open the perspective of optical information communication technologies [3].

The most typical applications in polarization-controlled devices contain three types of devices, namely directional coupling, focusing lens and plasmonic vortex lens. Directional coupling allows full control over the distribution of power between two counter-propagating SPP modes such as using meta-slit structures composed of arrays of subwavelength-scale nanoslit segments [3-5], U-shaped subwavelength plasmonic waveguides [6], single nanoslit [7], Δ-nanoantennas [8], split-ring-shaped slit resonators [9] and arrays of gap SPP resonators [10]. Focusing lens concentrates the SPPs to a designated focus, such as using nanoslits [11-14]. Plasmonic vortex lens (PVL) enables to focus the spin or orbital angular momentum to a specific spatial position in the form of a plasmonic vortex [15-25]. Unfortunately, most of the reported works realize these functions controlled only by the circularly or the linearly polarized light and this defect brings a lot of restrictions to enhance the device performances in the above areas.

Here, we propose a special method to realize the function of arbitrary spin-light directional coupling according to the input elliptically polarized light. More specifically, under the illumination of a left-handedness or a right-handedness light with the same ellipticity, the device is able to focus the light on the contrary side along a given direction. The direction of the excited SPPs can be designed according to our needs, and the incident polarization information has been preserved with the high extinction ratio simultaneously. It synthesizes the directional coupling and the near-field focus lens, which release a degree of freedom in the polarization-controlled devices and will be competent in the future.

## 2. Principle

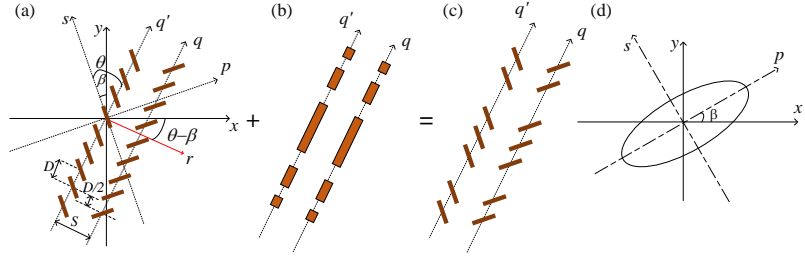

Fig. 1. (a) Schematic diagram of the directional coupling of SPPs. (b) The structure of Fresnel zone plate. (c) Schematic diagram of the directional focusing lens. (d) The input elliptical polarized light.

In the designed structure, there are two individual columns of air nanoslit arrays with a distance S equal to $\lambda_{spp}/4$ (or $-\lambda_{spp}/4$), as shown in Fig. 1(a). Because the SPPs emission pattern of the individual nanoslit is anisotropic, the excited SPPs' direction is along the $r$ or $-r$ which is perpendicular to the column direction [3]. The period in $q$-direction labeled by $D$ is $\lambda_{SPP}/2$. The offset of the columns is half a period ($D/2$), which help to reduce near-field coupling and scattering of the SPPs by neighboring apertures. $\beta$ is the angle between the long axis ($p$-axis) of the input elliptically polarized light with the $x$-axis, as shown in Fig. 1(d). $\theta$ denotes the direction of the nanoslit array with respect to slow axis ($s$-axis). The nanoslits in two columns are placed perpendicular to the long axis and slow axis respectively. The rectangular nanoslit arrays are fabricated in the 150-nanometer-thick gold surface and the dimensions of nanoslits are 50 nm * 200 nm. The input wavelength is 633 nm and the SPPs wavelength of $\lambda_{SPP}$ is calculated to be 606 nm. The focusing process can be understood with two steps. Let us consider the column in the Fig. 2(a) firstly. As we know, only the light with a polarization perpendicular to the nanoslits can be coupled into SPPs. The Jones matrix of input light is expressed by $[\cos\chi \quad \sigma i\sin\chi]^T$ in coordinate basis $[p \quad s]^T$, where $\chi \in (0, \pi/4)$ and $\sigma = \pm 1$ represent the left and right handedness respectively. Then the SPPs excited by the column pair can be expressed by

$$E_R = \cos\chi\cos\theta - \sigma i\sin\chi\sin\theta\exp(-i2\pi S/\lambda_{spp})$$
$$E_L = -\cos\chi\cos\theta + \sigma i\sin\chi\sin\theta\exp(i2\pi S/\lambda_{spp}) \qquad (1)$$

where $E_R$ and $E_L$ indicate the SPPs fields propagating to the right and the left sides. When $\theta = \pi/2 - \chi$, the light intensity of the opposite direction's SPPs is deduced as follows:

$$\begin{matrix} I_R = |\sin 2\chi|^2 \cos^2[\pi/4(1+\sigma)] \\ I_L = |\sin 2\chi|^2 \cos^2[\pi/4(1-\sigma)] \end{matrix} \text{ (or } \begin{matrix} I_R = |\sin 2\chi|^2 \cos^2[\pi/4(1-\sigma)] \\ I_L = |\sin 2\chi|^2 \cos^2[\pi/4(1+\sigma)] \end{matrix} \text{ when S}=-\lambda_{spp}/4\text{)}$$

For the left-handedness light, we have $I_R = 0$, $I_L = |\sin 2\chi|^2$, and it is inverse for the right-handedness light. The couple efficiency is $C \propto I_R + I_L = |\sin 2\chi|^2$ for both of them and the polarization extinction ratio (PER) is $\infty$ theoretically. However, the actual PER of the device will not reach $\infty$, since the nanoslits are discrete, and there are always amplitude and phase errors in the device, resulting in that the power in the right side is not null. More detail analyses will be shown later. Fig. 1 (b) shows the Fresnel zone plate based on the SPPs and (c) is the final schematic of the proposed structure which is able to realize the directional coupling and focusing.

## 3. Directional coupling of SPPs

We use a three-dimensional finite difference time domain (FDTD) method to obtain the SPPs field simulation results and a perfectly matched layer (PML) is utilized as the boundary condition. At first, we set $\beta=0$ and input different elliptical light of Gaussian beam. There are 11 pairs of nanoslit columns in the x direction and 25 nanoslits for each column and then we rotate the whole structure by $\theta$ ( $\theta=\pi/2-\chi$ )with respect to x clockwise for different input situation. The distance among these pairs is $\lambda_{SPP}$ and the SPPs excited by different pairs can interfere with each other constructively, which will help us get an obvious outcome. Light of different spin has been led to the opposite direction whose angle relative to the x-axis is $\theta$ respectively, as shown in the Fig. 2. The results turn out to be pretty good. However, it's difficult to calculate the couple efficiency and PER under this circumstance and the SPPs propagating along the x-axis can solve this problem. Also, it costs a lot of time to finish the simulation since there are many pairs. However, it can be seen that we can get an excellent result even if we employ only one pair of nanoslits in the following part.

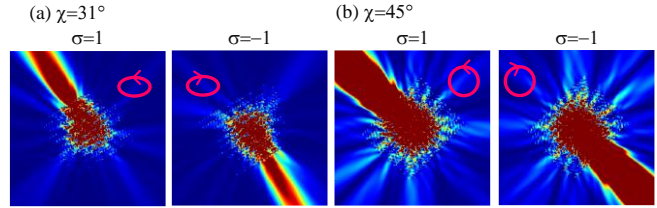

Fig. 2. The SPPs intensity distributions for various spin-controlled directional coupling when $\beta=0$ for $\chi=31°, \chi=45°$.

To attain unidirectional SPPs along x-direction in a shorter time, we set only one pair including 50 nanoslits whose schematic is exhibited in Fig. 3(a). This time we utilize planar wave as the input light with a series of ellipticities and the PML is also employed as the boundary condition. We need only to tune the nanoslits' angle for different inputs. From Fig. 3(b)-(d), one can see that our scheme can split various left-handedness and right-handedness elliptically polarized components to two contrary directions with a high PER, when the parameters of the device and the input light match each other. It's easy to see that the PER is getting higher gradually as ellipticity increases. And we can notice that even though the ellipticity is extremely low such as 11.3 degrees, the effect is still obvious.

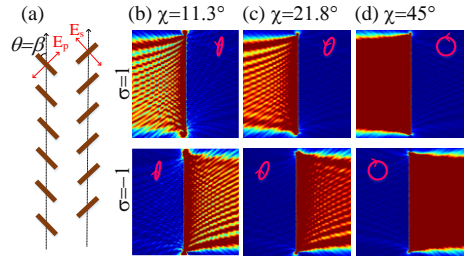

Fig. 3. (a)The schematic of the splitting part when β=θ and (b) The SPPs intensity distributions for various spin-controlled directional coupling.

The detail analyses have been done about the couple efficiency and PER. We put two linear monitor at $\pm 8.5 \mu m$ individually in x-axis. From Fig. 4, it's transparent to see that the couple efficiency is considerably close to the theoretical curve. As for the PER, we can know that the PER has almost overpassed 18dB as the ellipticity reach 21.8 degrees and it's impressively high compared to other similar devices that has been reported. The PER is low as the ellipticity is getting smaller, and it's in agreement with our prediction in the previous section. The excited SPPs' phase error of every column may cause this phenomenon. Here we define phase error as the SPPs' phase difference (subtracting $\pi/2$) excited by the separate column of every pair when propagating to the same cross section in z-axis from the original position. We set two linear polarization lights which are perpendicular to the nanoslit individually with a periodic boundary condition and obtain their phase difference. As we can see, the phase errors at low ellipticity is about 6 to 8 degrees whereas the error is only 2 degrees at high ellipticity. Because there is a larger initial phase difference for the low ellipticity, it inclines to mismatch the device parameter which leads to the low PER.

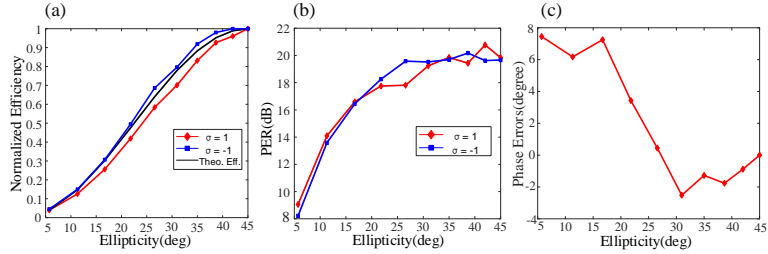

Fig. 4. (a) Normalized couple efficiency of different ellpticity where the circular polarization light's power has been set to 1. The red, blue, black line show the couple efficiency of splitting part for left, right-handedness and theoretical ones.; (b) Polarization extinction ratio for various spin-controlled directional coupling; (c) The SPPs' phase errors of individual column.

Then we increase to six pairs in x-direction to make the light-splitting effect better. As we can see in the Fig. 5(a), compared with degrees in Fig. 4 (b), the extinction ratio has enhanced obviously, from 9.06dB to13.39dB, 8.22dB to 14.87dB, for the left-handedness and the right one respectively. Moreover, when the ellipticity of input light is 2.86 degree, an excellent result that the PER are 8.86dB and 9.45dB for the opposite handedness light is still obtained. The promotion of coupling efficiency is caused by the increased coupling region. The increased PER is because the SPPs excited by each pair are constructively interfered in one direction.

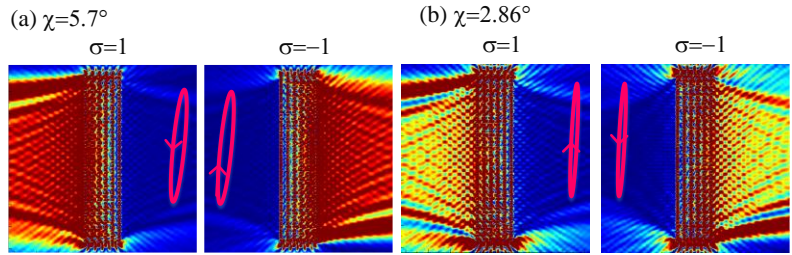

Fig. 5. (a)The power distribution of 6 pairs of columns and (b) The power distribution of input light whose ellpticity is 2.86 degrees.

## 4. Directional focusing of SPPs

Now we've got a device that is able to split the arbitrary polarization light of different spin perfectly. The next step of the focusing process is to realize a Fresnel zone plate based on the first part. We choose to block the even order zones and set the focal distance to $10\,\mu m$. The light excited by different odd order zones could interfere constructively and we could get a focus as we need. From the Fig. 7, it's clear to find out the input light is focused on a single point for different ellpticity, the contrary side of the structure.

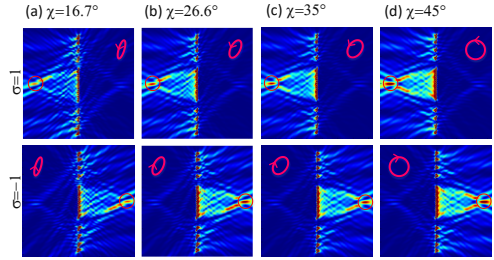

Fig. 6. The focus location for different elliptical light, 16.7, 26.6, 35, 45 degrees respectively.

The couple efficiency and the PER are shown in the Fig. 8. In this case, we choose the window function that covers the focus area when we calculate them above. The couple efficiency is still close to the theoretical result and the PER is pretty good in most situations. When the ellipticity is 5.7 degrees, the PER is extremely low especially for the right-handedness. The reason is due to the asymmetry of the structure in the y-direction which is used to improve the light intensity of propagating along the contrary orientation. Besides, since it's complicate to revise the window function according to every input light, we have to make it a little bigger than the precise area of the focus that introduce some stray light to the calculation. Of course, if one wants to enhance the effect of the directional coupling, we can add several pairs interfering constructively with each other.

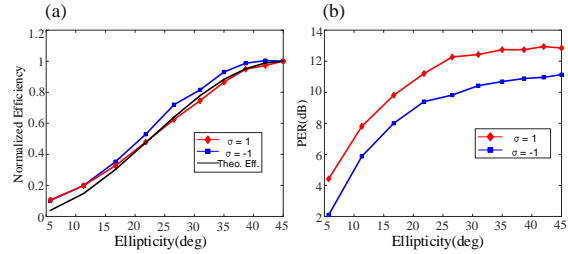

Fig. 7. (a) Couple efficiency of designed Fresnel lens. The red, blue and black line represents the same handedness of input as Fig.4. (b) Polarization extinction ratio for various spin-controlled directional focusing.

## 5. Conclusions

In summary, a new method is presented that is able to control the SPPs' directional coupling and focusing by inputting arbitrary spin of elliptical light with high PER. By tuning the angle of nanoslits of every columns and setting a specific distance between these columns, the device works very well according to our numerical achievements. We can understand the focusing process with two steps and the couple efficiency and the PER of every part has been discussed in detail. It is demonstrated that the characteristics of the device can remain stable in a large range of ellpticity. We expect that the proposed structure can be applied for polarization splitting, polarization sensing, polarization-multiplexed near-field images and surface plasmon holography in the future.

### Acknowledge

## References and links

1. W. L. Barnes, A. Dereux, and T. W. Ebbesen, "Surface plasmon subwavelength optics," Nature **424**, 824-830 (2003).
2. G. Rui and Q. Zhan, "Tailoring optical complex fields with nano-metallic surfaces," Nanophotonics **4**(2015).
3. J. Lin, J. P. Mueller, Q. Wang, G. Yuan, N. Antoniou, X. C. Yuan, and F. Capasso, "Polarization-controlled tunable directional coupling of surface plasmon polaritons," Science **340**, 331-334 (2013).
4. L. L. Huang, X. Z. Chen, B. F. Bai, Q. F. Tan, G. F. Jin, T. Zentgraf, and S. Zhang, "Helicity dependent directional surface plasmon polariton excitation using a metasurface with interfacial phase discontinuity," Light: Sci. Appl. **2**, e70 (2013).
5. J. P. Mueller, K. Leosson, and F. Capasso, "Polarization-selective coupling to long-range surface plasmon polariton waveguides," Nano Lett. **14**, 5524-5527 (2014).
6. Y. Lefier and T. Grosjean, "Unidirectional sub-diffraction waveguiding based on optical spin-orbit coupling in subwavelength plasmonic waveguides," Opt. Lett. **40**, 2890-2893 (2015).
7. S.-Y. Lee, H. Yun, Yohan, and B. Lee, "Switchable surface plasmon dichroic splitter modulated by optical polarization," Laser Photonics Rev. **8**, 777-784 (2014).
8. O. You, B. Bai, L. Sun, B. Shen, and Z. Zhu, "Versatile and tunable surface plasmon polariton excitation over a broad bandwidth with a simple metaline by external polarization modulation," Opt. Express **24**, 22061-22073 (2016).
9. Q. Xu, X. Zhang, Q. Yang, C. Tian, Y. Xu, J. Zhang, H. Zhao, Y. Li, C. Ouyang, Z. Tian, J. Gu, X. Zhang, J. Han, and W. Zhang, "Polarization-controlled asymmetric excitation of surface plasmons," Optica **4**, 1044 (2017).
10. A. Pors, M. G. Nielsen, T. Bernardin, J.-C. Weeber, and S. I. Bozhevolnyi, "Efficient unidirectional polarization-controlled excitation of surface plasmon polaritons," Light: Sci. Appl. **3**, e197 (2014).
11. S. Wang, X. Wang, Q. Kan, S. Qu, and Y. Zhang, "Circular polarization analyzer with polarization tunable focusing of surface plasmon polaritons," Appl. Phys. Lett. **107**, 243504 (2015).
12. T. Tanemura, K. C. Balram, D.-S. Ly-Gagnon, P. Wahl, J. S. White, M. L. Brongersma, and D. A. B. Miller, "Multiple-Wavelength Focusing of Surface Plasmons with a Nonperiodic Nanoslit Coupler," Nano Lett. **11**, 2693-2698 (2011).
13. S.-Y. Lee, K. Kim, S.-J. Kim, H. Park, K.-Y. Kim, and B. Lee, "Plasmonic meta-slit: shaping and controlling near-field focus," Optica **2**, 6-13 (2015).
14. G. H. Yuan, Q. Wang, P. S. Tan, J. Lin, and X. C. Yuan, "A dynamic plasmonic manipulation technique assisted by phase modulation of an incident optical vortex beam," Nanotechnology **23**, 385204 (2012).
15. G. Spektor, D. Kilbane, A. K. Mahro, B. Frank, S. Ristok, L. Gal, P. Kahl, D. Podbiel, S. Mathias, H. Giessen, F. J. Meyer Zu Heringdorf, M. Orenstein, and M. Aeschlimann, "Revealing the subfemtosecond dynamics of orbital angular momentum in nanoplasmonic vortices," Science **355**, 1187-1191 (2017).
16. Y. Gorodetski, A. Niv, V. Kleiner, and E. Hasman, "Observation of the Spin-Based Plasmonic Effect in Nanoscale Structures," Phys. Rev. Lett. **101**, 043903 (2008).
17. E. Ostrovsky, K. Cohen, S. Tsesses, B. Gjonaj, and G. Bartal, "Nanoscale control over optical singularities," Optica **5**, 283 (2018).
18. G. M. Lerman, A. Yanai, and U. Levy, "Demonstration of nanofocusing by the use of plasmonic lens illuminated with radially polarized light," Nano Lett. **9**, 2139-2143 (2009).
19. H. Kim, J. Park, S. W. Cho, S. Y. Lee, M. Kang, and B. Lee, "Synthesis and dynamic switching of surface plasmon vortices with plasmonic vortex lens," Nano Lett. **10**, 529-536 (2010).
20. W. Chen, D. C. Abeysinghe, R. L. Nelson, and Q. Zhan, "Experimental confirmation of miniature spiral plasmonic lens as a circular polarization analyzer," Nano Lett. **10**, 2075-2079 (2010).
21. W. Chen, D. C. Abeysinghe, R. L. Nelson, and Q. Zhan, "Plasmonic lens made of multiple concentric metallic rings under radially polarized illumination," Nano Lett. **9**, 4320-4325 (2009).
22. S. Yang, W. Chen, R. L. Nelson, and Q. Zhan, "Miniature circular polarization analyzer with spiral plasmonic lens," Opt. Lett. **34**, 3047-3049 (2009).
23. J. Li, C. Yang, H. Zhao, F. Lin, and X. Zhu, "Plasmonic focusing in spiral nanostructures under linearly polarized illumination," Opt. Express **22**, 16686-16693 (2014).
24. H. L. Zhou, J. J. Dong, Y. F. Zhou, J. H. Zhang, M. Liu, and X. L. Zhang, "Designing Appointed and Multiple Focuses With Plasmonic Vortex Lenses," IEEE Photon. J. **7**, 1-7 (2015).
25. H. Zhou, J. Dong, J. Zhang, and X. Zhang, "Retrieving orbital angular momentum distribution of light with plasmonic vortex lens," Sci. Rep. **6**, 27265 (2016).